\documentclass[pra,showpacs,showkeys,amsfonts,amsmath,twocolumn]{revtex4}
\usepackage{bm}
\usepackage{graphicx}
\RequirePackage{mathptm}

\numberwithin{equation}{section}
\newcommand{\si}[1]{\sigma_{#1}}
\newcommand{\sa}[2]{\sigma_{#1}^{#2}}

\newcommand{\W}[4]{\begin{cases}
#1 ,&#2\\
#3 ,&#4
\end{cases}}
\newcommand{\ro}{\varrho}
\newcommand{\ras}{\varrho_{\mr{as}}}
\newcommand{\la}{\lambda}

\newcommand{\ga}[1]{\gamma_{#1}}
\newcommand{\gh}{\widehat{\gamma}}

\newcommand{\Om}[1]{\Omega_{#1}}
\newcommand{\om}{\omega_{0}}
\newcommand{\oms}{\omega_{s}}

\newcommand{\I}{\openone}
\newcommand{\conj}[1]{\overline{#1}}
\newcommand{\ket}[1]{|{#1}\rangle}
\newcommand{\bra}[1]{\langle {#1} |}

\newcommand{\C}{\mathbb C}

\newcommand{\tr}{\mathrm{tr}\,}

\newcommand{\mr}[1]{\mathrm{#1}}

\newcommand{\rom}[1]{\rho_{#1}}

\newcommand{\msp}{\conj{M}}
\begin{document}
\title{Asymptotic entanglement of two atoms in squeezed light field}
\author{Lech Jak{\'o}bczyk
\footnote{ ljak@ift.uni.wroc.pl}, Robert Olkiewicz\footnote{ rolek@ift.uni.wroc.pl}
}
 \affiliation{Institute of Theoretical Physics\\ University of
Wroc{\l}aw\\
Plac Maxa Borna 9, 50-204 Wroc{\l}aw, Poland}
\author{Mariusz {\.Z}aba\footnote{ zaba@uni.opole.pl}}
\affiliation{Institute of Physics\\
Opole University\\
ul. Oleska 48, 45-052 Opole, Poland}
\begin{abstract}
The dynamics of entanglement between two - level atoms interacting with a common squeezed reservoir
is investigated. It is shown that for spatially separated atoms there is a unique asymptotic state
depending on the distance between the atoms and the atom - photons detuning. In the regime of strong correlations
there is a one - parameter family of asymptotic steady - states depending on initial conditions.
In contrast to the thermal reservoir both types of asymptotic states can be entangled.
We calculate the amount of entanglement  in the system in terms of concurrence.
\end{abstract}
\maketitle
\section{Introduction}
Dynamical creation of entanglement by the indirect interaction
between otherwise decoupled systems has been recently studied by
many researchers mainly in the case of two - level atoms interacting
with the common vacuum. The idea that dissipation can create rather
then destroy entanglement was put forward in several publications
\cite{Plenio, Kim, SM, PH}. In particular, the effect of spontaneous
emission on the destruction and production of entanglement was
discussed \cite{Ja, FT1,FT2,JJ1}. When the two atoms are separated by a small distance
compared to the radiation wavelength, there is a substantial probability that
a photon emitted by one atom will be absorbed by the other, and the resulting
process of photon exchange produces correlations between the atoms. Such correlations may cause
that initially separable states become entangled.
\par
The case of two atoms immersed in a common
thermal reservoir was also investigated \cite{BF,ZY,L,J}. As was shown in \cite{J}, similarly
to the vacuum case, the collective properties of the atomic system can alter the decay process
compared to the single atom. There are states with enhanced emission rates and such that the emission
rate is reduced. The important example of the latter is the antisymmetric superposition $\ket{a}$
constructed from energy levels of considered atoms. When the atoms are close to each other, this state
is decoupled from the environment and therefore is stable. In that case the asymptotic states of the system
are parametrized by the fidelity $F$ of the initial state with respect to the state $\ket{a}$ and the temperature
$T$ of the photon reservoir. Moreover, the asymptotic states can be identified with thermal generalization
of Werner states i.e. mixtures of the state $\ket{a}$ and Gibbs equilibrium state at the temperature $T$.
\par
In the present paper, we consider the atoms interacting with photon reservoir in a squeezed state \cite{LK}.
In practice,  squeezed light sources produce photon fields in multimode squeezed states, but here we assume
broadband approximation in which the parameters characterizing the photon field are constant over a sufficiently
broad frequency range. The dynamics of atoms interacting with squeezed light was studied by many authors (see e.g. review
paper \cite{DFS} and references therein). In the context of our studies, we mention the result of Palma and Knight
\cite{PK} showing the existence of highly correlated asymptotic state and the analysis of cooperative behavior
of atoms in broadband squeezed light in Ref. \cite{AP}.
\par
In this paper, we study the asymptotic entanglement of the system
of atoms evolving according the master equation considered by Tana{\'s} and Ficek \cite{TF} but we allow
non-zero detuning between the atomic transition frequency and the carrier frequency of the photon field. In the case
of spatially separated atoms studied in details in Ref. \cite{TF}, there exists a unique asymptotic state, but in
contrast to the vacuum or thermal reservoirs, this state can be entangled. However, the produced entanglement is
maximal only when the atoms are in resonance with the squeezed photon field. Non - zero detuning significantly
diminishes this production. When the atoms are close to each other, the dynamics of the system radically changes.
As in the vacuum or thermal case, the antisymmetric state again decouples from the reservoir, therefore is stable.
The asymptotic states $\ras$ depend on the initial fidelity $F$ and parameters describing the
reservoir, but non - zero detuning also modifies the matrix elements of $\ras$. We show that the asymptotic states
can be expressed as a mixture of a separable Gibbs state and two pure entangled states: an antisymmetric state $\ket{a}$
and some symmetric superposition of ground and excited levels of the atoms. This realization of the asymptotic state
simplifies for zero detuning and minimum - uncertainty squeezed reservoir to the mixture of $\ket{a}$ and
two - atom squeezed state \cite{PK}. Thus in that case, there are two linearly independent stable pure states so that
decoherence - free subspace is two - dimensional \cite{MO}.
\par
Depending on the initial fidelity, some of the asymptotic states are entangled.
We calculate the amount of the asymptotic entanglement using the concurrence as its measure. We also show that
for initial fidelity greater than some threshold value (depending on the properties of the reservoir and detuning),
the asymptotic concurrence is non - zero. This property is analogous to the thermal reservoir case.
But when the reservoir is in a squeezed state, somehow unexpected result occurs: initial states with small or even
zero fidelity become asymptotically entangled. The possibility of production of entanglement starting from separable
states with zero fidelity is very interesting. In this case the correlations present in a squeezed reservoir
are transferred to the atomic system, entangling for example two atoms
both in the ground state. But as before, large detuning between atoms and photon field, destroys this possibility.

\section{Model dynamics}
Consider two-level atoms $A$ and $B$ with ground states
$\ket{0}_{j}$ and excited states $\ket{1}_{j}$ ($j=A, B$),
interacting with the radiation field in a broadband squeezed vacuum
state with the carrier frequency $\oms$. The parameters $N$ and $M$
characterizing the squeezing satisfy
$$
M=|M|\,e^{i\vartheta}\quad\text{and}\quad |M|\leq \sqrt{N(N+1)},
$$
where the equality holds for a minimum-uncertainty squeezed state.
In the Markov approximation the influence of the reservoir on the
system of atoms can be described by the dynamical semi-group with
the Lindblad generator \cite{TF}
$$
L=-i[H,\cdot] +L_{\mr{D}},
$$
where
\begin{equation}
H=\frac{\om}{2}\sum\limits_{j=A,B}\sa{3}{j}+\sum\limits_{j,\,k=A,B\atop
j\neq k} \Om{jk}\sa{+}{j}\sa{-}{k},\label{hamilt}
\end{equation}
and
\begin{equation}
\begin{split}
L_{\mr{D}}\ro=&\frac{1}{2}\sum\limits_{j,\,k=A,\,B}\ga{jk}(1+N)\,\left(2\sa{-}{j}\ro\sa{+}{k}-\sa{+}{k}\sa{-}{j}\ro-\ro\sa{+}{k}\sa{-}{j}\right)\\
& +\frac{1}{2}\sum\limits_{j,\,k=A,\,B}\ga{jk}\,N\,\left(2\sa{+}{j}\ro\sa{-}{k}-\sa{-}{k}\sa{+}{j}\ro-\ro\sa{-}{k}\sa{+}{j}\right)\\
&+\frac{1}{2}\sum\limits_{j,\,k=A,\,B}\ga{jk}\,M\,\left(2\sa{+}{j}\ro\sa{+}{k}-\sa{+}{k}\sa{+}{j}\ro-\ro\sa{+}{k}\sa{+}{j}\right)\,e^{-2i\oms t}\\
&+\frac{1}{2}\sum\limits_{j,\,k=A,\,B}\ga{jk}\,\conj{M}\,\left(2\sa{-}{j}\ro\sa{-}{k}-\sa{-}{k}\sa{-}{j}\ro-\ro\sa{-}{k}\sa{-}{j}\right)\,e^{2i\oms
t}.
\end{split}\label{dyss}
\end{equation}
Here
$$
\sa{\pm}{A}=\sigma_{\pm}\otimes\I,\quad
\sa{\pm}{B}=\I\otimes\sigma_{\pm},\quad
\sa{3}{A}=\sigma_{3}\otimes\I,\quad \sa{3}{B}=\I\otimes\sigma_{3}.
$$
In the Hamiltonian (\ref{hamilt}), $\om$ is the frequency of the
transition $\ket{0}_{j}\to\ket{1}_{j}$ ($j=A,\, B$) and
$\Om{AB}=\Om{BA}=\Omega$ describes interatomic coupling by the
dipole-dipole interaction. On the other hand, dissipative dynamics is given by the generator (\ref{dyss}) with parameters
$\ga{AB}$ satisfying
\begin{equation}
\ga{AA}=\ga{BB}=\ga{0},\quad \ga{AB}=\ga{BA}=\gamma.\label{ga}
\end{equation}
In the above equalities, $\ga{0}$ is the single atom spontaneous
emission rate, and $\gamma=G(\vec{r}_{AB})\,\ga{0}$ is the collective damping
constant. In the model considered, $G(\vec{r}_{AB})$ is the function of the
interatomic distance $\vec{r}_{AB}$, and $G(\vec{r}_{AB})$ is small for large separation of
atoms. On the other hand, $G(\vec{r}_{AB})\to 1$ when $\vec{r}_{AB}$ is small (for more details see e.g.
\cite{FT}).
\par
The time evolution of the system of atoms is given by the master equation
\begin{equation}
\frac{d\ro}{dt}=L\,\ro, \label{me}
\end{equation}
In the frame rotating with frequency $\oms$, the master equation (\ref{me}) becomes an equation with time indendent
coefficients, and it may be written as
\begin{equation}
\frac{d\ro_{I}}{dt}=\widetilde{L}\,\ro_{I},\label{mer}
\end{equation}
where
$$
\widetilde{L}=-i[\widetilde{H},\cdot]+\widetilde{L}_{\mr{D}},
$$
with
\begin{equation}
\widetilde{H}=\frac{\delta_{0}}{2}\sum\limits_{j=A,B}\sa{3}{j}+\sum\limits_{j,\,k=A,B\atop
j\neq k} \Om{jk}\sa{+}{j}\sa{-}{k},\quad
\delta_{0}=\om-\oms,\label{hamiltr}
\end{equation}
and
\begin{equation}
\begin{split}
\widetilde{L}_{\mr{D}}\ro_{I}=&\frac{1}{2}\sum\limits_{j,\,k=A,\,B}\ga{jk}(1+N)\,\left(2\sa{-}{j}\ro_{I}\sa{+}{k}-\sa{+}{k}\sa{-}{j}\ro_{I}-
\ro_{I}\sa{+}{k}\sa{-}{j}\right)\\
& +\frac{1}{2}\sum\limits_{j,\,k=A,\,B}\ga{jk}\,N\,\left(2\sa{+}{j}\ro_{I}\sa{-}{k}-\sa{-}{k}\sa{+}{j}\ro_{I}-\ro_{I}\sa{-}{k}\sa{+}{j}\right)\\
&+\frac{1}{2}\sum\limits_{j,\,k=A,\,B}\ga{jk}\,M\,\left(2\sa{+}{j}\ro_{I}\sa{+}{k}-\sa{+}{k}\sa{+}{j}\ro_{I}-\ro_{I}\sa{+}{k}\sa{+}{j}\right)\\
&+\frac{1}{2}\sum\limits_{j,\,k=A,\,B}\ga{jk}\,\conj{M}\,\left(2\sa{-}{j}\ro_{I}\sa{-}{k}-\sa{-}{k}\sa{-}{j}\ro_{I}-\ro_{I}\sa{-}{k}\sa{-}{j}\right).
\end{split}\label{dyssr}
\end{equation}
Notice that in the Hamiltonian (\ref{hamiltr}), detuning $\delta_{0}$ can be arbitrary.
Only when the atoms are in resonance with the carrier frequency of the squeezed vacuum, $\delta_{0}=0$.
\par
From now on we omit the subscript $I$. The master equation
(\ref{mer}) can be used to obtain the equations for matrix elements
of a state $\ro$ of the system of two-level atoms with respect to
some basis. To simplify  the calculations one can work in the basis
of collective states in the Hilbert space $\C^{2}\otimes\C^{2}$
\cite{FT}, given by product vectors
\begin{equation}
\ket{e}=\ket{1}_{A}\otimes\ket{1}_{B},\quad
\ket{g}=\ket{0}_{A}\otimes\ket{0}_{B},
\end{equation}
symmetric superposition
\begin{equation}
\ket{s}=\frac{1}{\sqrt{2}}\left(\ket{0}_{A}\otimes\ket{1}_{B}+\ket{1}_{A}\otimes\ket{0}_{B}\right),
\end{equation}
and antisymmetric superposition
\begin{equation}
\ket{a}=\frac{1}{\sqrt{2}}\left(\ket{1}_{A}\otimes\ket{0}_{B}-\ket{0}_{A}\otimes\ket{1}_{B}\right).
\label{a}
\end{equation}
In the basis of collective states, two-atom system can be treated as
a single four-level system with ground state $\ket{g}$, excited
state $\ket{e}$ and two intermediate states $\ket{s}$ and $\ket{a}$. From (\ref{mer}) it follows that the matrix elements of the state
$\ro$ with respect to the basis $\ket{e},\, \ket{s},\,\ket{a},\,
\ket{g}$ satisfy the  equations which can be grouped into decoupled
systems of differential equations. So the diagonal matrix elements and $\rom{eg}$, satisfy
\begin{equation}
\begin{split}
\frac{d\rom{ee}}{dt}&=(\ga{0}-\gamma)N\,\rom{aa}+(\ga{0}+\gamma)N\,\rom{ss}-2\ga{0}N\,\rom{ee}\\
&\hspace*{4mm}-\gamma\,(M\rom{ge}+\msp\,\rom{eg}),\\
\frac{d\rom{ss}}{dt}&=-(\ga{0}+\gamma)[\,(1+2N)\,\rom{ss}-(1+n)\,\rom{ee}-N\,\rom{gg}\\
&\hspace*{4mm}-M\rom{ge}-\msp\,\rom{eg}\,],\\
\frac{d\rom{aa}}{dt}&=-(\ga{0}-\gamma)\,[\,(1+2N)\,\rom{aa}-(1+N)\,\rom{ee}-N\,\rom{gg}\\
&\hspace*{4mm}+M\rom{ge}+\msp\,\rom{eg}\,],\\
\frac{d\rom{gg}}{dt}&=(\ga{0}-\gamma)(1+N)\,\rom{aa}+(\ga{0}+\gamma)\,(1+N)\,\rom{ss}-2\ga{0}\,N\,\rom{gg}\\
&\hspace*{4mm} -\gamma\,(M\,\rom{ge}+\msp\,\rom{eg}),\\
\frac{d\rom{eg}}{dt}&=-(\ga{0}-\gamma)\rom{aa}+(\ga{0}+\gamma)\,M\,\rom{ss}-\gamma\,M\,\rom{gg}\\
&\hspace*{4mm}-(\ga{0}\,(1+2N)+2i\delta_{0})\,\rom{eg}.
\end{split}\label{diageg}
\end{equation}
On the other hand, the elements $\rom{ae},\, \rom{ag},\, \rom{se}$
and $\rom{sg}$ are connected by the following equations
\begin{equation}
\begin{split}
\frac{d\rom{ae}}{dt}&=\left[\,\gamma\,\left(N+\frac{1}{2}\right)-\ga{0}\left(2N+\frac{1}{2}\right)+i\,(\delta_{0}+\Omega)\right]\rom{ae}\\[2mm]
&\hspace*{4mm}-(\ga{0}-\gamma)\,\rom{ga}+(\ga{0}-\gamma)\msp\,\rom{ea}-\gamma\msp\,\rom{ag},\\[2mm]
\frac{d\rom{ag}}{dt}&=\left[\,\gamma\,\left(N+\frac{1}{2}\right)-\ga{0}\left(2N+\frac{1}{2}\right)-i\,(\delta_{0}-\Omega)\right]\rom{ag}\\[2mm]
&\hspace*{4mm}+(\ga{0}-\gamma)\,M\,\rom{ga}-(\ga{0}-\gamma)\,(1+N)\,\rom{ea}-\gamma\,M\,\rom{ae},\\[2mm]
\frac{d\rom{se}}{dt}&=-\left[\,\gamma\,\left(N+\frac{1}{2}\right)+\ga{0}\left(2N+\frac{1}{2}\right)-i\,(\delta_{0}-\Omega)\right]\,\rom{se}\\[2mm]
&\hspace*{4mm}+(\ga{0}+\gamma)\,\msp\,\rom{es}+(\ga{0}+\gamma)\, N\, \rom{gs}-\gamma\,\msp\, \rom{sg},\\[2mm]
\frac{d\rom{sg}}{dt}&=-\left[\,\gamma\,\left(N+\frac{1}{2}\right)+\ga{0}\left(2N+\frac{1}{2}\right)+i\,(\delta_{0}+\Omega)\right]\,\rom{se}\\[2mm]
&\hspace*{4mm}+(\ga{0}+\gamma)(1+N)\,\rom{es}+(\ga{0}+\gamma)\,M\,\rom{gs}-\gamma\,M\,\rom{se},
\end{split}\label{ae}
\end{equation}
and finally
\begin{equation}
\frac{d\rom{as}}{dt}=-\left[\ga{0}\,(1+2N)-2i\Omega\,\right]\,\rom{as}.\label{as}
\end{equation}
The equations for the remaining matrix elements can be obtained by using hermiticity of $\ro$.
\par
From the equations (\ref{diageg}) it follows that similarly as in the case of reservoir in the vacuum state (see e.g. \cite{FT}) and
thermal state \cite{J}, the system of atoms in the symmetric state $\ket{s}$ decays with the enhanced rate $\ga{0}+\gamma$, whereas
 antisymmetric initial state $\ket{a}$ leads to the reduced rate $\ga{0}-\gamma$. When the atoms are so close to each
 other that we can ignore the effects of their different spatial positions, we can put $\gamma=\ga{0}$. In this limiting case of strongly correlated atoms (Dicke model)
 the state $\ket{a}$ is completely decoupled from the reservoir. It can be also checked that the master equation
 (\ref{mer}) describes two types of time evolution of the system of atoms,
 depending on the relation between $\gamma$ and $\ga{0}$.
 When $\gamma <\ga{0}$, there is a unique asymptotic state. This
 state was found in Ref. \cite{TF} for the special case of zero
 detuning. In general case we compute it in the next section.
 On the other hand, in the Dicke model case when $\gamma=\ga{0}$, we show that there
 is a one - parameter family of
 nontrivial asymptotic states depending on the initial states.
\section{Asymptotic states}
\subsection{Spatially separated atoms}
We start with the case of spatially separated atoms, when
$\gamma<\ga{0}$. Direct calculations show that in that case, there
exists a unique stationary asymptotic state $\ro_{\mr{u}}$, which in the canonical
basis
$$
\ket{1}_{A}\otimes\ket{1}_{B},\;\ket{1}_{A}\otimes\ket{0}_{B},\;
\ket{0}_{A}\otimes\ket{1}_{B},\; \ket{0}_{A}\otimes\ket{0}_{B}
$$
has  non-vanishing matrix elements
\begin{equation}
\begin{split}
&\ro_{11}=\frac{a_{0}}{u_{0}},\quad \ro_{22}=\ro_{33}=\frac{c_{0}}{u_{0}},\\
& \ro_{23}=\frac{b_{0}}{u_{0}},\quad
\ro_{14}=\frac{z_{0}}{u_{0}},\quad \ro_{44}=\frac{d_{0}}{u_{0}},
\end{split}\label{rasg}
\end{equation}
where for
$$
\delta=\frac{\delta_{0}}{\ga{0}},\quad\gh=\frac{\gamma}{\ga{0}},
$$
we have
\begin{equation}
\begin{split}
u_{0}=&(1+2N)^{2}\,\left[\,(1+2N)^{2}+4\,\delta^{2}\,\right]\\
&+4|M|^{2}\,(\gh^{2}-(1+2N)^{2}),
\end{split}
\end{equation}
and
\begin{equation}
\begin{split}
a_{0}&=N^{2}\,\left[\,(1+2N)^{2}-4|M|^{2}+4\delta^{2}\right]+|M|^{2}\gh^{2},\\
c_{0}&=N(N+1)\,\left[(1+2N)^{2}-4|M|^{2}+4\delta^{2}\right]
+|M|^{2}\gh^{2},\\
d_{0}&=(1+N)^{2}\,\left[(1+2N)^{2}-4|M|^{2}+4\delta^{2}\right]+|M|^{2}\gh^{2}.
\end{split}\label{acdzero}
\end{equation}
Moreover,
\begin{equation}
\begin{split}
b_{0}&=-2\gh\,|M|^{2},\\
z_{0}&=-(1+2N-2i\delta)\,\gh\,M.
\end{split}\label{bzzero}
\end{equation}
The state (\ref{rasg}) in contrast to the analogous asymptotic state
in the thermal reservoir, can be entangled and, as we show later,
its entanglement crucially depends on the value of the normalized
detuning $\delta$ and the normalized
 damping constant $\gh$.
\subsection{Strongly correlated atoms}
When $\gamma=\ga{0}$, equations (\ref{diageg}) - (\ref{as}) simplify
and one can check that the solutions of (\ref{ae}) and (\ref{as})
asymptotically vanish and the only contribution to the asymptotic
states $\ras$ comes from $\rom{ee},\, \rom{aa},\, \rom{ss},\,
\rom{gg}$ and $\rom{eg}$. Note that in this case
$$
\frac{d\rom{aa}}{dt}=0,\quad\text{so}\quad
\rom{aa}(t)=\rom{aa}(0)=F,
$$
where
$$
F=\bra{a}\ro\ket{a}
$$
is the {\it fidelity} of the initial state $\ro$ with respect to the
antisymmetric state $\ket{a}$. Hence the fidelity the asymptotic state
$\ras$ also equals to $F$ and one finds that in the
canonical basis the matrix of $\ras$ has the same "X" form as in the
case of the state (\ref{rasg}), but with non - vanishing matrix
elements given by
\begin{equation}
\begin{split}
\ro_{11}&=(1-F)\,\frac{a}{u},\\
\ro_{22}&=(1-F)\,\frac{c}{2u}+\frac{F}{2},\\
\ro_{23}&=(1-F)\,\frac{c}{2u}-\frac{F}{2},\\
\ro_{14}&=(1-F)\,\frac{z}{u},\\
\ro_{44}&=(1-F)\,\frac{d}{u},
\end{split}\label{roasymp}
\end{equation}
and $\ro_{33}=\ro_{22}$. In the equations (\ref{roasymp}) we
have
\begin{equation}
\begin{split}
u=&(1+2N)^{2}\,(1+3N+3N^{2}-3|M|^{2})\\
&+4\,(1+3N+3N^{2})\,\delta^{2},
\end{split}\label{u}
\end{equation}
and
\begin{equation}
\begin{split}
a&=4N^{2}\,\left[N(N+1)-|M|^{2}\right]+|M|^{2}\\
&\hspace*{3mm}+N^{2}\,(1+4\delta^{2}),\\
c&=(1+2N)^{2}\,\left[N(N+1)-|M|^{2}\right]\\
&\hspace*{3mm} +2N(N+1)\delta^{2},\\
d&=(1+2N)\,\left[1+N+3(N(N+1)-|M|^{2})\right]\\
&\hspace*{3mm} +2N\,\left[N(N+1)-|M|^{2}\right]+4(1+N)^{2}\delta^{2}\\
z&=-(1+2N-2i\delta)\,M.
\end{split}\label{acdz}
\end{equation}
The asymptotic states $\ras$ defined by (\ref{roasymp}) exists for
any initial state, and for fixed parameters characterizing the
squeezing  depend on the initial fidelity and the normalized
detuning $\delta=\delta_{0}/\ga{0}$ of the electromagnetic field.
When $M=0$, we recover the case of standard thermal bath with $N$
playing the role of the mean photon number \cite{J}.
\par
To study the structure of the asymptotic states, we consider  first
the special case of a minimum-uncertainty squeezing and zero
detuning of the radiation field. One can check that in that case,
the matrix elements
 of $\ras$ are given by
\begin{equation}
\begin{split}
\ro_{11}&=(1-F)\,\frac{N}{1+2N},\\
\ro_{22}&=(\ras)_{33}=\frac{F}{2},\\
\ro_{23}&=-\frac{F}{2},\\
\ro_{44}&=(1-F)\,\frac{1+N}{1+2N},\\
\ro_{14}&=(1-F)\frac{\sqrt{N(N+1)}}{1+2N}\,e^{i\theta},
\end{split}\label{zerodet}
\end{equation}
where $\theta=\vartheta+\pi$. The asymptotic state given by
(\ref{zerodet}) has a remarkable structure: it is a mixture
\begin{equation}
\ras=(1-F)\,\ket{N,\theta}\bra{N,\theta}+F\,\ket{a}\bra{a}\label{raszerodet}
\end{equation}
of the pure state
\begin{equation}
\ket{N,\theta}=\sqrt{\frac{N}{1+2N}}\,\ket{0}_{A}\otimes\ket{0}_{B}+
e^{i\theta}\,\sqrt{\frac{1+N}{1+2N}}\,\ket{1}_{A}\otimes\ket{1}_{B}\label{Ntheta}
\end{equation}
and the antisymmetric state $\ket{a}$. The state $\ket{N,\theta}$ is known as two - atom
squeezed state, and can be obtained from the ground state $\ket{g}=\ket{0}_{A}\otimes\ket{0}_{B}$ by
applying the atomic squeezing transformation $S(\xi)$, given by
\begin{equation}
S(\xi)=\exp \,\left(\conj{\xi}\,\sa{-}{A}\sa{-}{B}-\xi\,\sa{+}{A}
\sa{+}{B}\right),
\end{equation}
for the appropriate choice of the complex parameter $\,\xi$
\cite{PK}. This state is entangled and in the limit of maximal
squeezing ($N\to\infty$), it becomes a maximally entangled
generalized Bell state. Notice also that $\ket{a}$ and
$\ket{N,\theta}$ span decoherence - free subspace for this specific
system, as it was recently established in Ref. \cite{MO}.
\par
In a general case the structure of $\ras$ is much more involved.
Define
\begin{equation}
F_{\mr{cr}}=\frac{c}{c+u}.\label{Fcr}
\end{equation}
By a direct calculation we see that if $F\geq F_{\mr{cr}}$, then
\begin{equation}
\ras=(1-p-q)\ro_{\beta}+p\,\ket{a}\bra{a}+q\,\ket{\psi}\bra{\psi},\label{rasgeneral}
\end{equation}
where
\begin{equation}
\begin{split}
p&=\left(1+\frac{c}{u}\right)\,F-\frac{c}{u},\\[2mm]
q&=\frac{|z|\,(a+d)}{u\,\sqrt{ad}}\,(1-F).
\end{split}\label{pq}
\end{equation}
The state $\ro_{\beta}$ is  a Gibbs state
\begin{equation}
\ro_{\beta}=\frac{e^{-\beta H_{\mr{a}}}}{\tr\, e^{-\beta
H_{\mr{a}}}}\label{robeta}
\end{equation}
for the Hamiltonian $H_{\mr{a}}=H_{0}+H_{1}$ with
$$
H_{0}=\frac{\om}{2}\sum\limits_{j=A,B}\sa{3}{j},\quad
H_{1}=\frac{\omega_{1}}{2}\left(\I\otimes\I+\sa{3}{A}\otimes\sa{3}{B}\right),
$$
the inverse temperature
\begin{equation}
\beta=\frac{1}{2\om}\,\ln\frac{d}{a},\label{beta}
\end{equation}
and the frequency
\begin{equation}
\omega_{1}=\frac{2\om}{\ln\,d/a}\,\ln\frac{c}{\sqrt{ad}-|z|}\,
.\label{omega1}
\end{equation}
Moreover, the pure state $\ket{\psi}$ is given by
\begin{equation}
\ket{\psi}=\sqrt{\frac{a}{a+d}}\:\ket{0}_{A}\otimes\ket{0}_{B} +
e^{i\phi}\,\sqrt{\frac{d}{a+d}}\:\ket{1}_{A}\otimes\ket{1}_{B},\label{psi}
\end{equation}
where $\phi=\arg\, z$.
\par
The formula (\ref{rasgeneral}) is a generalization of the equation
(\ref{raszerodet}) as well as the corresponding representation of
$\ras$ by the thermal generalization of Werner states in the case of
thermal reservoir \cite{J}. Observe also that for $F<F_{\mr{cr}}$,
the asymptotic state  cannot be expressed as the mixture
(\ref{rasgeneral}) but in contrast to the purely thermal case, the
states $\ras$ can be entangled even if $F<F_{\mr{cr}}$. We will
study this problem in the next section.
\section{Asymptotic entanglement}
For the characterization of entanglement of the asymptotic state
$\ras$ we use Wootters concurrence  \cite{Woot} defined for any
two-qubit state $\ro$ as
\begin{equation}
C(\ro)=\max
\left(\,0,\sqrt{\la_{1}}-\sqrt{\la_{2}}-\sqrt{\la_{3}}-\sqrt{\la_{4}}\,\right),
\end{equation}
where $\la_{1}>\la_{2}>\la_{3}>\la_{4}$ are the eigenvalues of the
matrix $\ro\widetilde{\ro}$ with $\widetilde{\ro}$ given by
$$
\widetilde{\ro}=\si{2}\otimes\si{2}\,\conj{\ro}\,\si{2}\otimes\si{2},
$$
where $\conj{\ro}$ denotes complex conjugation of the matrix $\ro$.
For the states in the "X" form, concurrence is given by the function
\begin{equation}
C(\ro)=\max \,\left(\,0,C_{1},\,C_{2}\,\right),\label{conc}
\end{equation}
with
\begin{equation}
\begin{split}
C_{1}&=2\,\left(\,|\ro_{14}|-\sqrt{\ro_{22}\ro_{33}}\,\right),\\
C_{2}&=2\,\left(\,|\ro_{23}|-\sqrt{\ro_{11}\ro_{44}}\,\right).
\end{split}
\end{equation}
\subsection{Entanglement of the asymptotic state $\ro_{\mr{u}}$}
Let us start with spatially separated atoms which have the unique
asymptotic state $\ro_{\mr{u}}$. Its concurrence is given by
\begin{equation}
C(\ro_{\mr{u}})=2\,\max\,\left(\,0,\frac{|z_{0}|-c_{0}}{u_{0}},\frac{|b_{0}|-\sqrt{a_{0}d_{0}}}{u_{0}}\,\right).
\label{konkrou}
\end{equation}
Analysis of this function  in the general case of broadband squeezed
reservoir is involved, so we focus on the case of minimum -
uncertainty squeezed states and consider (\ref{konkrou}) as the
function of squeezed field intensity $N$, for  fixed values of
parameters  $\gh$ and $\delta$. We plot this function in FIG.1 for
different values of detuning. It is evident that there is a range of
values of mean photon number $N$ for which the asymptotic
concurrence is positive. Observe that the maximum of
$C(\ro_{\mr{u}})$ appears for rather small values of $N$ and the non
- zero detuning diminishes the production of entanglement.
\begin{figure}[h]
\centering
{\includegraphics[height=54mm]{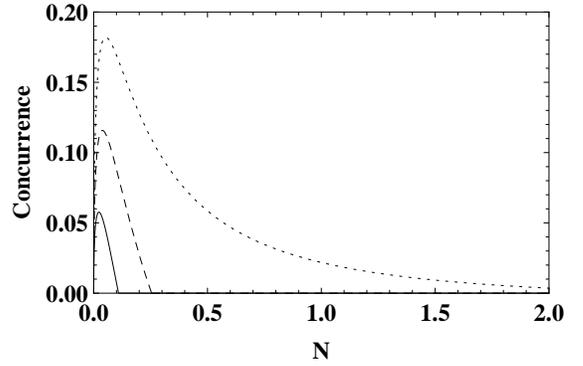}}\caption{Entanglement of the state $\ro_{\mr{u}}$ as a function of N for $\gh=0.85$
and different values of detuning: $\delta=0$ (dotted curve); $\delta=0.5$ (dashed curve) and $\delta=1$ (solid curve).}
\end{figure}
\subsection{Entanglement of the states $\ras$}
The properties of the concurrence of $\ras$ as a function of initial
fidelity can be studied in more details. Notice that for  these
states we have
\begin{equation}
C_{1}=\left(\frac{c-2|z|}{u}-1\,\right)\,F-\frac{c-2|z|}{u}.\label{c1}
\end{equation}
Define
\begin{equation}
F_{1}=\max\,\left(0,\,\frac{c-2|z|}{c-2|z|-u}\,\right).\label{F1}
\end{equation}
If $F_{1}>0$, then
$$
C_{1}>0\quad\text{for}\quad 0\leq F<F_{1}.
$$
On the other hand
\begin{equation}
C_{2}=2\,\left(\left|(1-F)\,\frac{c}{2u}-\frac{F}{2}\right|-(1-F)\frac{\sqrt{ad}}{u}\,\right).
\end{equation}
Notice that if $F<F_{\mr{cr}}$, then
$$
(1-F)\,\frac{c}{2u}-\frac{F}{2}>0,
$$
and
\begin{equation}
C_{2}=\left(\frac{2\sqrt{ad}-c}{u}-1\,\right)\,
F-\frac{2\sqrt{ad}-c}{u}.
\end{equation}
Since
$$
\frac{2\sqrt{ad}-c}{u}-1<0,
$$
so
$$
C_{2}<0\quad\text{when}\quad F<F_{\mr{cr}}.
$$
\par
 Let now $F\geq F_{\mr{cr}}$, then
\begin{equation}
C_{2}=\left(1+\frac{c+2\sqrt{ad}}{u}\,\right)\,F-\frac{c+2\sqrt{ad}}{u}.\label{c2}
\end{equation}
Define
\begin{equation}
F_{2}=\frac{c+2\sqrt{ad}}{c+2\sqrt{ad}+u}\label{F2}
\end{equation}
Form the equation (\ref{c2}) we see that
$$
C_{2}>0\quad\text{when}\quad F>F_{2}.
$$
On the other hand, direct calculations show that
$$
F_{2}\geq F_{\mr{cr}}\quad\text{and}\quad F_{1}\leq F_{2}.
$$
 \begin{figure}[b]
\centering
{\includegraphics[height=54mm]{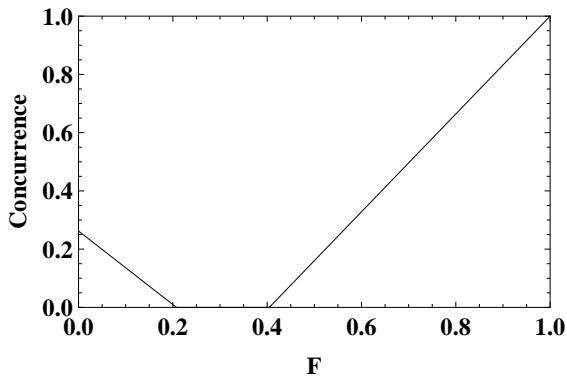}}\caption{Asymptotic entanglement versus fidelity for minimum - uncertainty squeezed reservoir with $N=1$ and detuning $\delta =0.8$}
\end{figure}
Taking into account the above results we arrive at the conclusion
that depending on the initial fidelity $F$, the asymptotic state
$\ras$ is entangled for all $F\in [0,F_{1})\cup (F_{2},1]$ (provided
$F_{1}>0$) and separable for $F\in [F_{1},F_{2}]$ (see FIG.2). The asymptotic
concurrence reads
\begin{equation}
C(\ras)=\W{C_{1}}{0\leq F<F_{1}}{C_{2}}{F_{2}<F\leq 1}
\end{equation}
with $C_{1}$ and $C_{2}$ given by equations (\ref{c1}) and
(\ref{c2}), respectively. This general result covers also the
special cases of vacuum reservoir where $F_{2}=0$, thermal reservoir
with $F_{1}=0$ and $F_{2}>0$ and minimum-uncertainty squeezed
reservoir, where $F_{1}=F_{2}$. It is worth to stress that the
creation of the asymptotic states with non-zero entanglement from
the initial states with small  or even zero fidelity is only
possible when the reservoir is in a squeezed state. Let us discuss
this point in more details in a special case of atoms which are in
resonance with minimum - uncertainty radiation field. In this case
\begin{equation}
F_{1}=F_{2}=\frac{2\sqrt{N(N+1)}}{2\sqrt{N(N+1)}+(1+2N)},
\end{equation}
and
\begin{equation}
C(\ras)=\W{-(1+C_{0})F+C_{0}}{F<F_{1}}{\hspace*{3mm}(1+C_{0})F-C_{0}}{F>F_{1}}
\end{equation}
with
\begin{equation}
C_{0}=2\frac{\sqrt{N(N+1)}}{1+2N}.\label{c0}
\end{equation}
 \begin{figure}[b]
\centering
{\includegraphics[height=54mm]{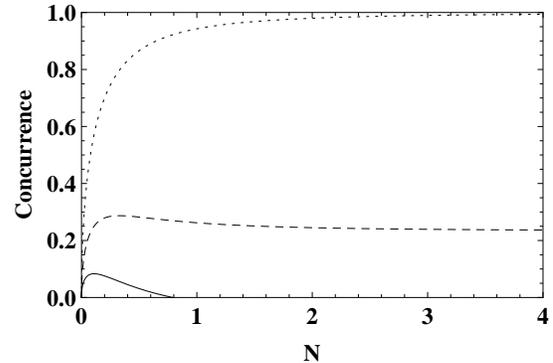}}\caption{Asymptotic entanglement of initial states with $F=0$
as a function of $N$, for different values of detuning:
$\delta=0$ (dotted curve); $\delta=0.8$ (dashed curve) and $\delta=2$ (solid curve)}
\end{figure}
For all initial states with zero fidelity, we obtain pure entangled
state (\ref{Ntheta}) with concurrence equal to $C_{0}$. Notice that
in the limit of maximal squeezing, this state becomes maximally
entangled. For pure product states
\begin{equation}
\ket{\Psi}=\ket{\varphi}\otimes\ket{\psi}
\end{equation}
the fidelity is given by the formula
\begin{equation}
F=\frac{1}{2}\,\left(1-|\langle\varphi\ket{\psi}|^{2}\,\right),
\end{equation}
so the zero fidelity corresponds, for example, to the case of two
atoms prepared in the same initial states. This leads to the
remarkable result: the interaction with squeezed reservoir will
entangle two atoms which are initially in the ground state
$\ket{g}=\ket{0}_{A}\otimes\ket{0}_{B}$. The analogous phenomenon
cannot occur when the photon field is in the vacuum or thermal
state. Notice also that for non - zero detuning, the asymptotic
state is no longer pure
 and the production of stationary entanglement is less effective (FIG.3).
\section{Conclusions}
We have investigated the dynamics of two - level atoms interacting with the photon reservoir
in a broadband squeezed vacuum state. Time evolution of the system crucially depend on the relative distance
between the atoms. When the atoms are spatially separated, there is a unique asymptotic state,
which can be entangled in contrast to the analogous asymptotic state for the thermal reservoir.
In the case of small interatomic distance, there are nontrivial asymptotic states $\ras$ which are
parametrized by the fidelity $F$
and the parameters $N$ and $M$ characterizing the squeezing. The states $\ras$ depend also on the detuning
between atomic transition frequency and carrier frequency of the photon field. For the values of $F$ above
the threshold fidelity $F_{2}$, the states $\ras$ are entangled. Non - zero entanglement can also occur for
small values  of $F$ or even if $F=0$, and this possibility is a unique feature of the squeezed reservoir. When
the atoms are in resonance with the photon field and $|M|=\sqrt{N(N+1)}$, the asymptotic state corresponding to
$F=0$ is the pure entangled state known as two - atom squeezed state.

\end{document}